\documentclass[letterpaper, inpress]{jds} 
\usepackage[table]{xcolor}
\usepackage{url}
\usepackage{hyperref}
\usepackage{booktabs}
\usepackage{lipsum}
\usepackage{comment}
\usepackage{array}
\usepackage{graphicx}
\usepackage{subcaption}
\usepackage{mwe}
\usepackage{float}
\usepackage{caption}
\usepackage{lscape}
\usepackage{bm}
\usepackage{svg}
\usepackage{url}
\usepackage{ulem}
\hypersetup{
  colorlinks   = true, 
  urlcolor     = blue, 
  linkcolor    = blue, 
  citecolor   = blue 
}

\jdsmonth{October}                 
\jdsyear{2022}                  
\jdsvolume{xx}                  
\jdsissue{xx}                   
\jdsdoi{xx.xxxx/xxxxxxxxx}      
\jdsreceived{October, 2022}       
\shortauthors{Amona E, et al.}

\usepackage{amsfonts,amsmath,amssymb,amsthm}
\usepackage{booktabs}

\usepackage{lipsum}

\title[Incorporating Interventions to a SEIRDV Model]{Incorporating Interventions to an Extended SEIRD Model with Vaccination: Application to COVID-19 in Qatar}

\author[1]{Elizabeth B Amona}
\author[2]{Ryad A Ghanam}
\author[1]{Edward L Boone}
\author[1]{Indranil Sahoo \thanks{Corresponding author. Email: sahooi@vcu.edu}}
\author[3]{Laith J Abu-Raddad}
\affil[1]{Department of Statistical Sciences and Operations Research, Virginia Commonwealth University, Richmond, Virginia, 23284, USA}
\affil[2]{Department of Liberal Arts and Sciences, Virginia Commonwealth University in Qatar, Education City, Doha, Qatar}
\affil[3]{Weill Cornell Medicine, Qatar}
\begin{document}

\maketitle

\begin{abstract}
  The Covid-19 outbreak of 2020 has required many governments to develop and adopt mathematical-statistical models of the pandemic for policy and planning purposes. To this end, this work provides a tutorial on building a compartmental model using Susceptible, Exposed, Infected, Recovered, Deaths and Vaccinated (SEIRDV) status through time. The proposed model uses interventions to quantify the impact of various government attempts made to slow the spread of the virus. Furthermore, a vaccination parameter is also incorporated in the model, which is inactive until the time the vaccine is deployed. A Bayesian framework is utilized to perform both parameter estimation and prediction. Predictions are made to determine when the peak Active Infections occur. We provide inferential frameworks for assessing the effects of government interventions on the dynamic progression of the pandemic, including the impact of vaccination. The proposed model also allows for quantification of number of excess deaths averted over the study period due to vaccination. 

\end{abstract}


\section{Introduction}
\label{sec:intro}
\noindent In 2019, the Coronavirus Disease (COVID-19) \citep{wu2020new, rezabakhsh2020novel} appeared in the human population in Wuhan, China, and a pandemic ensued affecting the entire world. It spread rapidly through 196 countries, requiring strict precautions to attempt to control it's spread.  Governments set policies (interventions) such as wearing masks, hand sanitizing, and social distancing to mitigate the progression of the pandemic \citep{giuliani2020modelling}.  Early in the pandemic, the State of Qatar was one of the middle eastern countries that had high numbers of residents diagnosed to have COVID-19 with the first infected case on February 29, 2020 which quickly spread in the population with $362,007$ confirmed cases by April 5, 2022 \citep{chinazzi2020effect}.  In the subsequent months, stricter policies were enacted to slow the progression while researchers looked for cures and treatments.  Several vaccines were developed to slow the spread of the disease, however, the models in the mathematical and statistical literature have not been directly applied to data to discern the impact of vaccines.

\cite{Bertsimas2020optimizing} proposed extending the traditional Susceptible, Exposed, Infected, Recovered (SEIR) models which focused on governmental policies, response of the society, and reduced mortality rates. They also proposed a ``DELPHI-V'' model which captures the effects of vaccinations and examined the impact of COVID-19 on mortality. This paper focused on optimizing vaccine allocations and simulating the pandemic dynamics using a coordinate descent algorithm.  Another study incorporated a vaccine compartment into a SIRD model focusing on the vaccine allocation to the susceptible individuals using an optimization approach based on Thompson sampling (TS) to understand vaccine efficiency mean rates over time \citep{Rey2021vaccine}. Furthermore, \citet{Ghostine2021extended} extended a SEIR model with a vaccine compartment, and implemented the ensemble Kalman filter (EnKF) to improve the forecasting ability of their model. This work also examined the effect of vaccination on the spread of COVID-19. Another study by \citet{Wintachai2021stability} attempted to understand the effectiveness of prophylactic and therapeutic vaccines by observing the reproduction number before the introduction of the vaccine and how the reproduction number curve flattens after the introduction of the vaccine using numerical simulations. However, this study did not fit their model to the data to obtain the parameter estimates but relied solely on the literature to obtain these coefficients. 

Notice that all the above-mentioned studies did not directly incorporate intervention measures enacted by the government into their model, and therefore, could not explain the impact of vaccines on the transmission rates as the government policies were deployed. Furthermore, in models where coefficients are determined from previous literature, it is difficult to ensure that all uncertainties are adequately quantified. In addition, a recent study by \citet{Koufi2020dynamics} examined the dynamics of the SIRS epidemic model under the switching of regimes, which is related to the idea of this work, however, no data was used for the analysis and their model did not incorporate the compartment that determines where the regime-switching will occur. Lastly, prior to the development of vaccines, \citet{ghanam2021seird, boone2021monitoring} developed a Susceptible, Exposed, Infected, Recovered and Deaths (SEIRD) model that incorporated interventions to understand the impact of changes in government policies on the spread of COVID-19.  While this modeling approach was able to quantify the early interventions by the government, it will not able to determine the impact of the introduction of vaccines into the population and the resulting impacts on disease transmission. Furthermore, the model proposed did not develop the associated Reproduction number.

To address the inadequacies of the previous approaches, this work develops a Susceptible, Exposed, Infected, Recovered, Death and Vaccinated (SEIRDV) model using an intervention paradigm similar to \citet{ghanam2021seird} with the inclusion of the associated reproduction number. Here, another compartment (Vaccination) is added to explain the switching of regimes before and after the addition of this compartment. Simply adding a compartment for Vaccine is not sufficient as it does not include the possibility of vaccine inefficacy. Also, \citet{ghanam2021seird} assumed that the transmission rates after each intervention are dependent, however, this paper assume that these transmission rates are not dependent, therefore, we study the impact of interventions by observing these coefficients independently.

This work is organized in the following manner. In Section~\ref{sec:SEIRDV} the Susceptible, Exposed, Infected, Recovered, Death and Vaccinated (SEIRDV) model that is employed is defined. 
Section~\ref{sec:Data} describes the data available for the State of Qatar. Section \ref{sec:Inter} shows how interventions are incorporated into the model. The Bayesian inference model specification is given in Section~\ref{sec:Bayes}. Section~\ref{sec:Results} summarizes the results, including the estimated time varying reproduction and the projected number of deaths without vaccination. Finally, Section~\ref{sec:Discussion} provides a discussion of the method and some insights into implementing the method for policy making.

\section{The SEIRDV Model}\label{sec:SEIRDV}
\noindent Recall, the SIR model \citep{kermack1927contribution} transitions individuals between Susceptible, Infected and Recovered compartments.  Additional compartments such as Exposed (asymptomatic), Death, Quarantine, Hospitalized, and Vaccinated can be added to better model the dynamics of an infectious disease, as it moves through the population.  Furthermore, the model should reflect the reality of policies enacted by the associated government. The data illustration presented in this paper focuses specifically on the COVID-19 pandemic in the State of Qatar and the corresponding policies implemented during the study time. Based on the availability of data, several key assumptions are made:
\begin{enumerate}
  \item Emigration and birth rate are excluded from the model.
  \item Once a person is in the infected (symptomatic) compartment, they are quarantined and hence do not interact with the susceptible population, except for caregivers who contract the disease at a separate rate.
  \item The recovered and deaths compartments are for those who are first infected.
  \item Only individuals who were exposed and recovered after being exposed get the vaccine, while individuals who were infected and recovered do not get vaccinated since they already develop antibodies. 
  \item The susceptible, exposed and recovered individuals become vaccinated at the same rate.
  \item Once vaccinated, the individual remains insusceptible to reinfection during the study time.
\end{enumerate}

Details of the model compartments are given in Table \ref{tbl:compart}, parameter descriptions and associated units are presented in Table \ref{tbl:Param}.

\large{
\begin{table}[ht!]
\caption{Description of compartments considered in model \ref{eq:Sys1}}\label{tbl:compart}
\begin{center}
\begin{tabular}{l l } \hline

Compartment & Description \\
\hline
$S(t)$ & Number of susceptible individuals  at time $t$  \\
$E(t)$  & Number of exposed individuals at time $t$  \\
 $I(t)$  & Total number of infected individuals at time $t$ \\
 $R_E(t)$ & Cumulative number of the exposed individuals who recovered at time $t$ \\
 $R_I(t)$ & Cumulative number of the infected individuals who recovered at time $t$ \\
$D(t)$ & Cumulative number of deaths at time $t$   \\
  $V(t)$  & Number of vaccinated individuals at time $t$. \\
\hline
\end{tabular}
\end{center}
\end{table}}

\begin{table}[ht!]
\caption{Explanation of parameters considered in model \ref{eq:Sys1}, along with the associated units}\label{tbl:Param}
\begin{center}
\scalebox{0.9}{
\begin{tabular}{l l l} \hline
Parameter & Description & unit \\
\hline
$\alpha$ & Transmission rate from Susceptible to Exposed & per day$\times$individual$^2$  \\
$\beta$ & Rate at which Exposed become Infected & per day \\
 $\gamma$  & Rate at which both the Exposed and Infected become Recovered & per day \\
 $\zeta$ & Mortality rate for those Infected & per day  \\
$\rho$ & Vaccination rate for Susceptible, Exposed  and Recovered from exposed & per day  \\
\hline
\end{tabular}}
\end{center}
\end{table}


 \begin{figure}[ht!]
\begin{center}
\includegraphics[width = 0.6\textwidth]{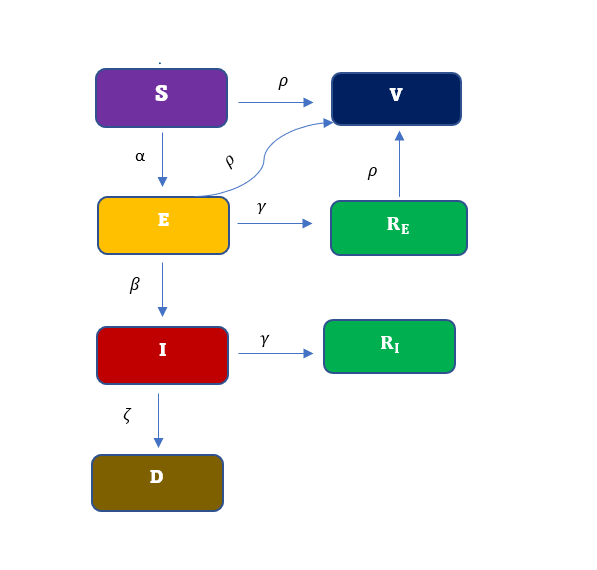} 
\caption{Schematic diagram of SEIRDV model for Covid-19 . \label{fig:Schematic}}
\end{center}
\end{figure}

The compartmental model proposed, contains compartments for Susceptible, Exposed, Infected, Recovered from Exposed state, Recovered from Infected state, Deaths and Vaccinated.  The schematic diagram for this novel SEIRDV model is shown in Figure~\ref{fig:Schematic}. Notice that a Susceptible individual, $S$ becomes Exposed (asymptomatic), $E$ at rate $\alpha$ and the Exposed (asymptomatic) will either recover, $R_E$ at rate $\gamma$ or become sick and move to the Infected compartment, $I$ at rate $\beta$. The Infected can recover, $R_I$ at rate $\gamma$, or die, $D$ at rate $\zeta$. Lastly, the Susceptible, Exposed and the Exposed who recover can get Vaccinated at the same rate $\rho$. Based on model assumptions, individuals who recover after being infected develop antibodies and do not require vaccination during the study period. Also, once an individual is vaccinated, the person remains in the compartment.

Mathematically, the model in Figure \ref{fig:Schematic} is expressed as the following system of first-order nonlinear ordinary differential equations:

\begin{equation}
\label{eq:Sys1}
\begin{cases}
\frac{dS}{dt} = -\alpha S(t) E(t) - \rho S(t) \\
\frac{dE}{dt} = \alpha S(t) E(t) - (\beta + \gamma + \rho) E(t) \\
\frac{dI}{dt} =  \beta E(t)  - (\gamma + \zeta)I(t) \\
\frac{dR_E}{dt} = \gamma E(t) - \rho R_E(t)\\
\frac{dR_I}{dt} = \gamma I(t) \\
\frac{dD}{dt} = \zeta I(t) \\
\frac{dV}{dt} = \rho S(t) +  \rho E(t) + \rho R_E(t),
\end{cases}
\end{equation}
with the following constraints $~S(t) \geq 0, ~ E(t) \geq 0, ~I(t) \geq 0, ~ R_E(t) \geq 0, ~ R_I(t) \geq 0, ~D(t) \geq 0,~ \text{and} ~V(t) \geq 0$.  Notice this formulation has the Vaccine compartment in the model during a time-frame for which no vaccine is available.  By specifying $\rho= 0$ and $V(0) = 0$, the model collapses to a SEIRD model and will exhibit the associated dynamics.  Once the vaccine is deployed, the restriction $\rho > 0$ is employed in order to ``activate'' the Vaccine compartment where the value of $\rho$ reflects the rate at which people who are susceptible, exposed and recovered after being exposed, are being vaccinated. This approach allows the parameter estimates to be found using all the data with no need to switch models.  

\subsection{Model Analysis}
\subsubsection{Effective Reproduction Number and Disease-Free Equilibrium}
\noindent The term ``disease free'' means that there is no disease in the system, hence no one is infected and everyone is assumed susceptible, although there are a few exposed individuals. Note that the vaccine compartment has been activated at this point, so $\rho > 0$. Therefore, estimating the basic reproduction number $(R_0)$ might not be a good idea since we are no longer in a fully susceptible population. A better metric that would capture the introduction of vaccination is the effective reproduction number ($ \mathfrak{R_e}=R_0 (S/N)$) \citep{Mercer2011effective, Ridenhour2014unraveling}. Thus, this work considers the effective reproduction number, which also takes into account  the interventions enacted by the government in addition to the vaccine compartment.

The disease-free equilibrium of the system in (\ref{eq:Sys1}), denoted by $X^{0}$ is:

\[X^{0} = \left( S^{0} \approx  N, E^{0} = 5, I^{0}= 0, R_E^{0}= 0, R_I^{0}= 0, D^{0}= 0, V^{0}=0 \right) \]

Note the $I^{0}, R_E^{0}, R_I^{0}, D^{0}$ and $V^{0}$ were directly obtained from the data. $S^{0} = S(0) \approx N$, where $N$ is the total population since we assumed that everyone is susceptible before the introduction of vaccine. 

The reproduction number, $R_{0}$ is computed using the next-generation matrix approach \citep{van2002reproduction}. Let $X = (E,I)^T$, then,  the system of equations in (\ref{eq:Sys1}) can be written as
\[\dfrac{dX}{dt} = G(X)-W(X),\] 
where $G(X) =(\alpha S E , 0)^T$ and $W(X) = ((\beta + \gamma + \rho)E - \beta E, ( \gamma + \zeta )I)^T$. The corresponding disease-free Jacobian matrices of $G(X)$ and $W(X)$ are:
\[J(G(X)) = G = \begin{bmatrix}
\alpha S^{0}  &   0\\
0        &  0
\end{bmatrix}.   \]

\[J(W(X)) = W = \begin{bmatrix}
(\beta + \gamma + \rho ) & 0 \\
-\beta       &  ( \gamma + \zeta)
\end{bmatrix}   \]
According to \cite{van2002reproduction}, $R_{0}$ is defined as the spectral radius of the next-generation matrix $(GW^{-1})$. Hence, we need to compute $GW^{-1}$, and its spectral radius. Now,

\[ W^{-1} = \begin{bmatrix}
\frac{1}{(\beta + \gamma + \rho)}  & 0\\ \vspace{0.5cm}
\frac{\beta}{(\beta + \gamma + \rho)( \gamma + \zeta )}  & \frac{1}{(\gamma + \zeta)}.
\end{bmatrix}\]
Therefore,

\[ G W^{-1} = \begin{bmatrix}
\alpha S^{0} & 0 \\
0        &  0
\end{bmatrix}  \begin{bmatrix}
\frac{1}{(\beta + \gamma + \rho)}  & 0\\ \vspace{0.5cm}
\frac{\beta}{(\beta + \gamma + \rho)( \gamma + \zeta )}  & \frac{1}{(\gamma + \zeta)}.
\end{bmatrix}.
\]

\[= \begin{bmatrix}
\frac{\alpha S^{0} }{(\beta + \gamma + \rho)}  & 0\\
0         & 0
\end{bmatrix} \]

At the zeroth generation where $S^{0}$ remains the total population, the above expression becomes

\[GW^{-1} = \begin{bmatrix}
\frac{\alpha S^{0} }{ (\beta + \gamma + \rho)} & 0\\
0         & 0
\end{bmatrix}. \]

\noindent Now, the spectral radius of a matrix $A$, denoted by $\mathcal{R}(A)$ is defined as the largest modulus of the eigenvalues of $A$. Thus, the basic reproduction number for this model is given by
\begin{eqnarray*}\label{eq:R0}
R_{0} = \mathcal{R}(GW^{-1}) =  \frac{\alpha S^{0}}{ (\beta + \gamma + \rho)}.
\end{eqnarray*}
Also, $ \mathfrak{R_e} =  R_0 (S(0)/N)$. Since we assumed that $S(0) \approx N$, the two terms will cancel out, which means that the effective reproduction number is approximately equal to the basic reproduction number, (i.e., $\mathfrak{R_e} \approx  R_0$). Therefore, the effective reproduction number is given by
\begin{eqnarray}\label{eq:Re}
\mathfrak{R_e} = \frac{\alpha S^{0}}{ (\beta + \gamma + \rho)},
\end{eqnarray}
a metric used to measure the transmission potential of a disease in the presence of an intervention. It is defined as the average number of new infections produced by an infectious individual in a population. Thus, we expect an endemic state when $ \mathfrak{R_e} > 1$, and a disease-free state when $ \mathfrak{R_e} < 1$  \citep{van2002reproduction, Ridenhour2014unraveling}. That is, we expect an endemic state if $\alpha  S^{0} > (\beta + \gamma + \rho)$ and a declining state if $\alpha S^{0} < (\beta + \gamma + \rho)$. Notice that in this case, ($\mathfrak{R_e} \approx   R_0$) despite the introduction of vaccination and intervention measure. A plausible reason for this is our model assumption that the total population of Qatar are susceptible at the start of the epidemic. However, we expect this value to decrease with time and $ \mathfrak{R_e} \neq  R_0$, especially when vaccine is introduced in the later part of the study time. 


\section{Data Description}\label{sec:Data}
The Johns Hopkins University (JHU) Covid-19 Github site \citep{miller20202019} includes  daily cumulative number of confirmed infections, cumulative number of recovered and the cumulative number of deaths for every country starting January 22, 2020. All data for Qatar during the period of study was obtained from this site.  Notice that in model (\ref{eq:Sys1}) the Recovered and Death states are cumulative as once one enters these compartments, there is no exit.  However, the Infected compartment has transitions from Exposed and to Recovered and Deaths.  Hence the data for confirmed infections are cumulative and include both Recovered and Deaths.  As such, if $CI(t)$ be the number of confirmed infections as reported by JHU at time $t$, then the number of infected subjects at time $t$ is defined as
\begin{equation*}
I(t) = CI(t) - R(t) - D(t).
\end{equation*}
\noindent For clarity, the term ``Active Infections'' will be used to denote the derived variable, $I(t)$, versus the Cumulative Infected, $CI(t)$, provided in the data. Note that information on vaccinated individuals became available for the State of Qatar on 29th of April, 2021.

\begin{figure}
 \begin{subfigure}{0.5\textwidth}
     \includegraphics[width=\textwidth]{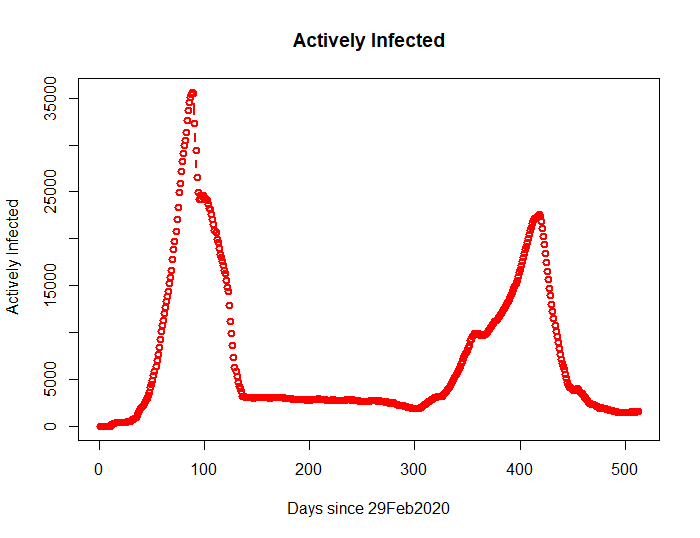}
     \caption{Number of active infections }
     \label{fig:a1}
 \end{subfigure}
 \hfill
 \begin{subfigure}{0.5\textwidth}
     \includegraphics[width=\textwidth]{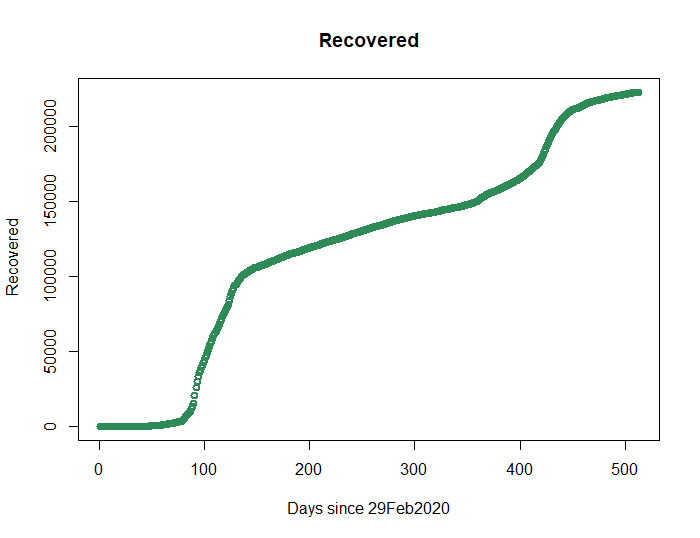}
     \caption{Cumulative number of recovered individuals}
     \label{fig:b1}
 \end{subfigure}
 
 \medskip
 \begin{subfigure}{0.5\textwidth}
     \includegraphics[width=\textwidth]{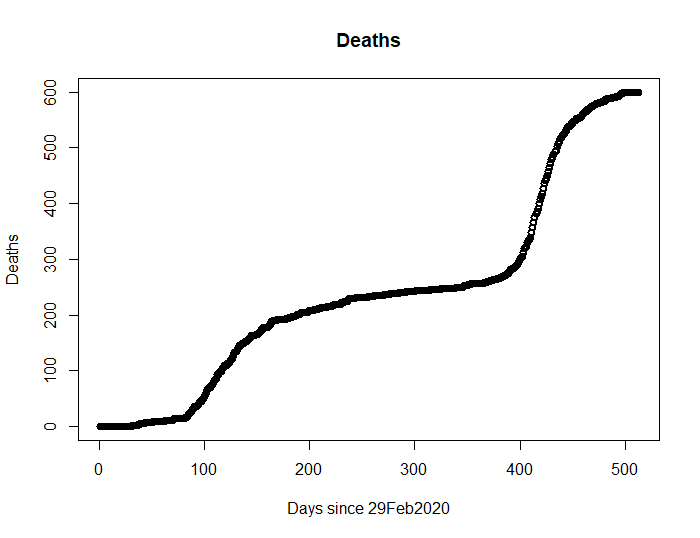}
     \caption{Cumulative number of deaths}
     \label{fig:c1}
 \end{subfigure}
 \hfill
 \begin{subfigure}{0.5\textwidth}
     \includegraphics[width=\textwidth]{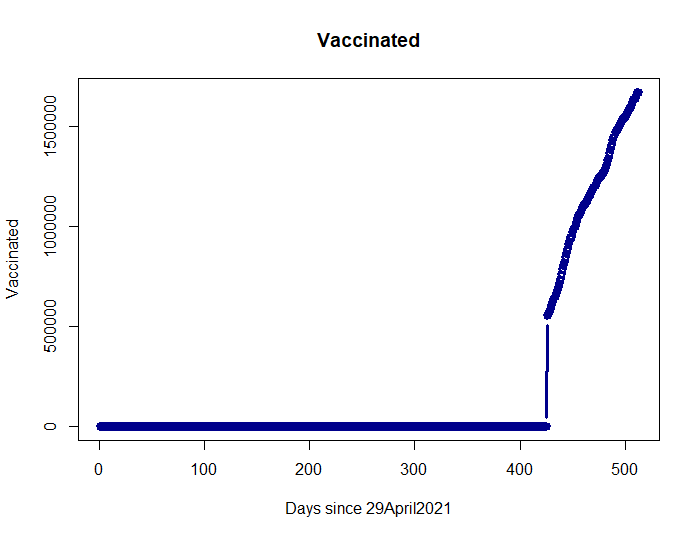}
     \caption{Cumulative number of vaccinated individuals}
     \label{fig:d1}
 \end{subfigure}

 \caption{Plots of (a) Active Infections, (b) Recovered, (c) Deaths (February 29, 2020 - October 13, 2021) and (d) Vaccinated (April 29, 2021 - October 13, 2021) for the State of Qatar. }
 \label{fig:InitialData}

\end{figure}

Figure~\ref{fig:InitialData} shows the plots of daily Active Infections, Recovered and Deaths data for the State of Qatar since February 29, 2020, and the number of vaccinated individuals since April 29, 2021.  The active infections are very low until around day 35 when there is large jump due to increase in testing. The active infections then plateaus until day 300, after which there is another extreme growth.  There seems to be a similar pattern for the number of recovered individuals with a delay showing the time of infection before recovery.  The plot for Deaths shows no deaths until day 95 and then a steady increase in deaths for the remaining days. Finally, the plot for Vaccinated shows that vaccination information became available on day 426 (April 29, 2021) and has increased steadily since then. 

\section{Incorporating Interventions}\label{sec:Inter}
\noindent Despite the various intervention measures taken by the Qatari government, there have been a large number of infected cases in Qatar. It is quite plausible that some interventions were more helpful than others, and some were not helpful at all. For example, on day 48 (March 10, 2020), the Qatari government announced the closure of all schools and universities due to COVID-19 outbreak, and placed a travel ban on 15 countries. On day 54, (March 15, 2020), the government included three additional countries to its travel ban. This is documented in the Qatar National Preparedness and Response Plan for Communicable Diseases published by the Ministry of Public Health, Qatar and Major Risks to Business Continuity published by the Hamad Medical Corporation. The Ministry of Municipal and Environment on day 60 (March 21, 2020), ordered the closing of all parks and public beaches and enacted a policy where both public and private sector employees must conduct 80\% of their work from home. On day 71 (April 1, 2020), the Council of Ministers and Industry decided to temporarily close all restaurants, cafes, and food trucks. On day 78 (April 8, 2020), the Ministry of Commerce mandated that people work from home to curb the spread of coronavirus. On day 93 (April 23, 2020), the holy month of Ramadan began where people were not allowed the opportunity to mingle and celebrate with others. On day 115 (June 4, 2020), the cabinet decided to allow only four people in a vehicle that could accommodate more people under normal circumstances and permitted working hours for only the private sector from 7 am until 8 pm. However, as documented in the Qatar National Preparedness and Response Plan for
Communicable Diseases published by the Ministry of Public Health, Qatar and Major Risks to Business Continuity published by the Hamad Medical Corporation, restrictions placed by the ministry of commerce and industry were lifted between June 15 and September, 2020. Therefore, it is of interest to understand the impact of government interventions on the transmission rate. Do these interventions increase, decrease or have no impact on the transmission rate? Section \ref{sec:Results} provides answer to this question.
\vspace{0.25mm}

Since the government regulations can directly influence the rate at which the susceptible population become exposed ($\alpha$), the transmission rate, more emphasis will be given on this parameter. The proposed method incorporates this idea using indicator functions, denoted by $\Phi_m(t)$ defined as
\begin{equation*}
\Phi_m(t) = \begin{cases} 1 & \text{if } t > t_m \cr
0 & \text{otherwise}
\end{cases}
\end{equation*}
\noindent where $t_m$ is the time where the $m^{th}$ intervention occurs and index $m = 1,2,..,M$.  For each intervention, there needs to be a change in the value of $\alpha$, denoted by $\alpha_m$, that captures the impact of the intervention.  Let $\bm{\Phi}(t) = \left( 1, \Phi_1(t), \Phi_2(t), ... , \Phi_m(t) \right)^{T}$ be the vector of values of each $\Phi_m(t)$ at time $t$. Also, let $\bm{\alpha} = \left( \alpha_0, \alpha_1,...\alpha_m \right)^{T}$, where each of the $\alpha's$ are independent, which means that we do not need $\alpha_i$ to obtain $\alpha_{i+1}$. Thus, the transition rates between $S(t)$ and $E(t)$ are given by
\begin{equation*}
\alpha(t) = \begin{cases} \alpha_0 & \text{if } 0 < t_1< t \cr
 \alpha_1 & \text{if } 0 < t_2< t \cr
 \alpha_2 & \text{if } 0 < t_3< t \cr
\vdots & \vdots \cr
\alpha_m & \text{if } 0 < t_m< t.
\end{cases}
\end{equation*}
\noindent Since $\alpha(t) > 0$ for all $t$, the following constraints are required
\begin{eqnarray*}
\alpha_0 &>& 0   \cr
 \alpha_1 &>& 0  \cr
\alpha_2 &>& 0  \cr
\vdots & >& \vdots  \cr
\alpha_m &>& 0.
\end{eqnarray*}

Since the recovery rate could also be influenced by the intervention measures, we define $\bm{\gamma}= (\gamma_0, \gamma_1, \cdots, \gamma_n)^T, n \leq m$, where each of the $\gamma's$ are independent and denotes the changed recovery rate once an intervention has been administered. Since we assume that exposed and infected individuals recover at the same rate, the transition rates between $I(t)$ and $R_I(t)$ and $E(t)$ and $R_E(t)$ are given by

\begin{equation*}
\gamma(t) = \begin{cases} \gamma_0 & \text{if } 0 < t_1< t \cr
 \gamma_1 & \text{if } 0 < t_2< t \cr
 \gamma_2 & \text{if } 0 < t_3< t \cr
\vdots & \vdots \cr
\gamma_n & \text{if } 0 < t_n< t.
\end{cases}
\end{equation*}
\noindent Since $\gamma(t) > 0$ for all $t$, the following constraints are required
\begin{eqnarray*}
\gamma_0 &>& 0   \cr
 \gamma_1 &>& 0  \cr
\gamma_2 &>& 0  \cr
\vdots & >& \vdots  \cr
\gamma_n &>& 0.
\end{eqnarray*}

Furthermore, impulse functions can be used to model the spike (dramatic shift) when transitioning between states.  This can be represented as a Dirac delta, defined by

\begin{equation*}
\delta (x) = \begin{cases} +\infty, & \text{~if~ } x =0 \cr
0, & \text{~if~ } x\neq 0,
\end{cases}
\end{equation*}
\noindent which satisfies $\int_{-\infty}^{\infty} \delta(x) dx = 1$ \citep{dirac1958general}. This can be integrated into the model to capture spikes in the number of infected individuals.  In our application, the State of Qatar data exhibits this type of behavior at day 35 when one can clearly see a large jump in the number of infections. This is incorporated into the model using the Dirac delta function, $\delta(t-\tau)$. The impact of the jump (spike) is captured in the model by examining the transmission rate between Exposed and Infected with a coefficient $\beta^{*}$.

Thus, the proposed SEIRDV model after incorporating interventions is given by 

\begin{equation}
\label{eq:inten}
\begin{cases}
\frac{dS}{dt} = -\bm{\alpha} S(t) E(t) - \rho S(t) \\
\frac{dE}{dt} = \bm{\alpha} S(t) E(t) - (\beta + \bm{\gamma} + \rho) E(t) - \beta^{*} E(t)\delta(t-\tau) \\
\frac{dI}{dt} =  \beta E(t)  - (\bm{\gamma} + \zeta)I(t)+ \beta^{*} E(t) \delta(t-\tau) \\
\frac{dR_E}{dt} = \bm{\gamma} E(t) - \rho R_E(t)\\
\frac{dR_I}{dt} = \bm{\gamma} I(t) \\
\frac{dD}{dt} = \zeta I(t) \\
\frac{dV}{dt} = \rho S(t) +  \rho E(t) + \rho R_E(t),
\end{cases}
\end{equation}
The proposed model in Equation \ref{eq:inten} now takes into account interventions administered by the government to better model their impact on the dynamics of the pandemic. 

\subsection{Time Varying Effective Reproduction Number}

\noindent The time varying effective reproduction number, $\mathfrak{R_e (t)}$ is now defined as the average number of people an infectious person will infect at time $t$. In this study, we consider the time varying effective reproduction number, which means that we are interested in knowing the number of secondary cases an infectious person can produce throughout the period of infection. Since the the number of susceptible, vaccinated and the parameters in (\ref{eq:R0}) are time varying, we can write our time varying effective reproduction number as follows:
\begin{eqnarray}\label{eq:R0(t)}
\mathfrak{R_e}(t) = \frac{\alpha_i S^{0}(t)}{(\beta + \gamma_j + \rho)},
\end{eqnarray}
where $\alpha_i$ and $\gamma_j$ are the $i^{th}$ and $j^{th}$ components of $\bm{\alpha}$ and $\bm{\gamma}$ respectively ($i=0,1, 2,\cdots,m, ~ j =0,1,2\cdots,n$). From equation \ref{eq:R0(t)}, we see that $\mathfrak{R_e}(t)$ changes when the transmission and recovery rates change, thereby showing the impact of interventions on the effective reproduction number.


\section{Statistical Methodology}\label{sec:Bayes}
\noindent Bayesian framework is used for this analysis due to the complexity of the model. The traditional Bayes' formula \citep{bayes1763lii} is defined as

\begin{equation}\
\pi (\theta | \mathbf{D} ) = \frac{ \pi( \theta )L( \mathbf{D} | \theta ) }{ \int_{\Theta} \pi( \theta )L( \mathbf{D} | \theta ) d\theta} \nonumber
\end{equation}
where $\pi( \theta | \mathbf{D})$ is the posterior probability distribution for the parameters $\theta$ given the data $\mathbf{D}$, $\pi(\theta)$ is the prior distribution of $\theta$ and $L( \mathbf{D} | \theta )$ is the likelihood of the data given $\theta$.

\noindent To obtain the likelihood of the model in equation (\ref{eq:inten}), the {\it mean} abundance of each compartment in the model is considered and is given by
\begin{eqnarray}\label{eq:Sys2}
\frac{d \phi_S(t)}{d t} &=&  -\Phi(t)^T\bm{ \alpha } \phi_S(t)\phi_E(t) - \rho \phi_S(t) \cr
\frac{d \phi_E(t)}{d t} &=& \Phi(t)^T\bm{ \alpha } \phi_S(t)\phi_E(t) - (\beta + \bm{\gamma} + \rho) \phi_E(t)  - \beta^{*} \phi_E(t)\delta(t-\tau)  \cr
\frac{d \phi_I(t)}{d t} &=& \beta \phi_E(t)  - \bm{\gamma} \phi_I(t) - \zeta \phi_I(t)  + \beta^{*} \phi_E(t)\delta(t-\tau)  \cr
\frac{d \phi_{R_E}(t)}{d t} &=& \bm{\gamma} \phi_E(t) - \rho \phi_{R_E}(t) \cr
\frac{d \phi_{R_I}(t)}{d t} &=& \bm{\gamma} \phi_I(t) \cr
\frac{d \phi_D(t)}{d t} &=& \zeta \phi_I(t) \cr
\frac{d \phi_V(t)}{d t} &=& \rho\phi_S(t) + \rho \phi_E(t) + \rho \phi_{R_E}(t).
\end{eqnarray}

\noindent where $\phi_S(t)$, $\phi_E(t)$, $\phi_I(t)$, $\phi_{R_E}(t)$, $\phi_{R_I}(t)$, $\phi_D(t)$ and $\phi_V(t)$ are the means of $S(t)$, $E(t)$, $I(t)$, $R_E(t)$, $R_I(t)$, $D(t)$ and $V(t)$ respectively and the parameters have the same definition as provided in the system given in equation (\ref{eq:Sys1}). Since there is no data for $S(t)$, $E(t)$ and $R_E(t)$, these compartments will be latent variables and will not be directly included in the likelihood. The likelihoods for $I(t)$, $R_I(t)$, $D(t)$ and $V(t)$ are given by

\begin{eqnarray}\label{eq:Like1}
I(t) &\sim& Poisson \left( \phi_I(t) \right) \cr
R_I(t) &\sim& Poisson \left( \phi_{R_I}(t) \right) \cr
D(t) &\sim& Poisson \left( \phi_D(t) \right) \cr
V(t) &\sim& Poisson \left(\phi_V(t) \right)
\end{eqnarray}

To specify the prior distributions for $\bm{ \alpha }$, $\beta^{*}$, $\beta$, $\bm{\gamma}$, $\zeta$ and $\rho$ the following constraints are necessary $\alpha > 0$, $\beta > 0$, $\beta^{*} \ge 0$ $\gamma > 0$, $\zeta > 0$ and $\rho \ge 0$.  The prior distribution for $\rho$ gives the researcher the ability to ``activate'' the Vaccine compartment in the model at time $T_V$ by specifying a point mass prior, $\rho = 0$, for the time period before the deployment and $Exp(1)$ once the vaccine is deployed. Therefore, the prior distributions are as follows
\begin{eqnarray}
{ \alpha_i } &\sim& Exp(1), \hspace{0.25cm} i = 0, 1, \cdots, m \cr
\beta^{*} &\sim& Exp(1) \cr
\beta &\sim& Exp(1) \cr
\gamma_j &\sim& Exp(1), \hspace{0.25cm} i = 0, 1, \cdots, n \cr
\zeta &\sim& Exp(1) \cr
\rho &\sim& \begin{cases} P(\rho = 0) = 1 & t < T_V \\
                            Exp(1) & t \ge T_V 
                            \end{cases} 
\end{eqnarray}

The posterior distribution obtained from the likelihood and prior distribution specification is given by
\begin{eqnarray*}
 \pi \left( \bm{ \alpha }, \beta^{*}, \beta, \bm{\gamma}, \zeta, \rho| \mathbf{D} \right)  &\propto& \scriptstyle \pi(\bm{ \alpha }) \pi( \beta^{*}) \pi(\beta)\pi(\bm{\gamma})\pi(\zeta) \pi(\rho) L(\mathbf{D}|\bm{ \alpha }, \beta^{*}, \beta,\bm{\gamma},\zeta,\rho) \cr
&\propto&  e^{-\sum\limits_{i}\alpha_i  - \beta^{*} - \beta - \sum\limits_{j}\gamma_j -\zeta-\rho }    \cr
&\times& \prod_{t=1}^{T}\frac{\phi_I(t)^{I(t)} \phi_{R_I}(t)^{R_I(t)}\phi_D(t)^{D(t)}\phi_V(t)^{V(t)}e^{-\phi_I(t)- \phi_{R_I}(t) - \phi_D(t)-\phi_V(t)}}{I(t)!R_I(t)!D(t)!V(t)!}
\end{eqnarray*}
Since the posterior distribution does not have an analytical solution, Markov chain Monte Carlo (MCMC) techniques were used to sample from the posterior distribution \citep{gelman1995bayesian}.  Specifically Metropolis-Hastings sampler was used to obtain samples from the posterior distribution \citep{gilks1995markov, albert2009introduction, ghanam2021seird}.  To tune the sampler, a series of short chains were generated and analyzed for convergence and adequate acceptance rates.  These initial short chains were discarded as ``burn-in'' samples.  The tuned sampler was used to generate 30,000 samples from $\pi( \bm{ \alpha }, \beta^{*}, \beta, \bm{\gamma}, \zeta, \rho| \mathbf{D})$ and trace plots were visually examined for convergence.  All inferences were made from these 30,000 posterior samples. The model and sampling algorithm was custom programmed in the R statistical programming language version 3.6.3. The computations take approximately 2400 seconds using an AMD A10-9700 3.50GHz processor with 16GB of RAM to obtain 30,000 samples from the posterior distribution. For more details on statistical inference see \citet{wackerly2014mathematical, casella2021statistical, berger1985prior}.


\section{Results}\label{sec:Results}
The initial conditions used for the analysis are specified in Table \ref{tbl:Init}. $S(0)$ was chosen as the population of the State of Qatar as of April 1, 2022, $I(0), ~R_E(0)=R_I(0),  ~D(0)$ and $V(0)$ were obtained directly from the data. $E(0)$ was chosen based on sensitivity analysis. It is worth mentioning that values of $E(0)$ between 5 and 10 fit the model well and had the same Psuedo $R^{2} \approx 99.9\%$. $E(0)=5$ was chosen as this value isn't too large to cause a fast spread of the disease. Furthermore, model interventions were placed at days $t_1 = 12,~t_2 = 35,~t_3 = 48,~t_4 = 60,~t_5 = 71,~t_6 = 78,~t_7 = 87,~t_8 = 93,~t_9 = 104,~t_{10} = 115,~t_{11} = 136,~t_{12}= 350,~t_{13} = 355$ and $t_{14} = 420 ~ (m = 14)$ with a Dirac delta impulse at time $\tau = 35$.

\begin{table}[ht!]
\caption{The initial conditions for the model }\label{tbl:Init}
\begin{center}
\begin{tabular}{l c c c } \hline

Compartment & Initial conditions \\
\hline
$S(0)$ & $2,782,000$ & Population of Qatar as of April $1$, 2022 \\
$E(0)$  & $5$ & Assumed\\
 $I(0)$  & 1 & From the data (JHU) \\
 $R_E(0) = R_I(0) = 0$ & 0 & From the data (JHU)\\
$D(0)$ & 0 & From the data (JHU) \\
  $V(0)$  & 0  & From the data (JHU) \\
\hline
\end{tabular}
\end{center}
\end{table}

Table~\ref{tbl:ParEst} shows the posterior means, standard deviations and the 0.025\%, 0.5\% and 0.975\% quantiles for the model parameters based on the 30,000 samples from the posterior distribution.  Notice that, $\widehat{\alpha}_0 = 1.42\times 10^{-7}$ and $\widehat{\alpha}_1 = 4.62\times 10^{-8}$ are quite close, indicating that the first intervention resulted in a low transmission rate.  Similarly we can see that the second and third interventions $\widehat{\alpha}_2 = 5.99 \times 10^{-8}$ and $\widehat{\alpha}_3 = 4.38 \times 10^{-8}$ have some overlap resulting in a very low transmission rate. Furthermore, $\widehat{\alpha}_4= 4.32 \times 10^{-8}$ is a low decrease as well as $\widehat{\alpha}_5 = 3.63 \times 10^{-8}$ which  resulted in a low transmission rate. However, $\widehat{\alpha}_6= 4.96 \times 10^{-8}$ is a moderate increase with another moderate increase in $\widehat{\alpha}_7= 8.24 \times 10^{-8}$. Also, $\widehat{\alpha}_8= 5.52 \times 10^{-8}$ is a moderate decrease, $\widehat{\alpha}_{9}= 9.63 \times 10^{-12}$ is a moderate increase, $\widehat{\alpha}_{10}= 5.20 \times 10^{-8}$ is a moderate decrease, $\widehat{\alpha}_{11}= 5.79 \times 10^{-8}$ is a moderate increase, but $\widehat{\alpha}_{12}= 1.13 \times 10^{-8}$ is a very low decrease, while $\widehat{\alpha}_{13}= 3.35 \times 10^{-8}$ and the final transmission rate $\widehat{\alpha}_{14}= 8.16 \times 10^{-8}$ both have a moderate increase. It is interesting to see that the estimated mortality rate $\widehat{\zeta} = 0.000121 \approx 1/8264$ which means that about 1 in 8,264 people dies from the disease each day, which is quite low.  Also, the estimated infection rate is $\widehat{\beta} = 0.05386 \approx 1/18.57$, which corresponds to about 1 in $18.57$ exposed people become infected each day. The quantile intervals provide a 95\% credible intervals for the parameters and can be used to obtain a range of reasonable parameter values. For example for the parameter $\beta$ the interval is  (0.05350, 0.05420) meaning the probability that $\beta$ is between (0.05350, 0.05420) is 0.95. This translates to an interval for risk interpretations as approximately $1/0.05420 \approx 18.45$ to $1/0.05350 \approx 18.69$ people are exposed to the infection every day. This also gives insight into the number of exposed individuals in the population who may be infected but do not yet exhibit symptoms. The rates at which people recover from the disease at different time points during the study time, $\gamma_0=0.00846, \gamma_1=0.01323, \gamma_2=0.10092, \gamma_3=0.09407, \gamma_4=0.07121, \gamma=0.01979$ and $\gamma_6=0.09309$ are all estimated to be between $0.00846$ (min) which correspond to $\gamma_0$ and $0.10092$ (max) which correspond to $\gamma_2$. This tells us that approximately $1$ person recovers from the disease each day. This value may be low due to delays in reporting. However, this value seems reasonable, considering the total population of Qatar.

\begin{table}[ht!]
\caption{Posterior Mean, Median, Standard Deviation and ($Q_{0.025},Q_{0.5},Q_{0.975}$) for $\alpha_0$, $\alpha_1$, $\alpha_2, \cdots, \alpha_{14}$, $\beta*$, $\beta$, $\gamma_0, \gamma_1, \cdots, \gamma_6$, $\zeta$, and $\rho$. The posterior estimates are based on 30,000 samples from the posterior distribution.}\label{tbl:ParEst}
\centering
\scalebox{0.9}{
\begin{tabular}{|lcccc|} \hline
Parameter & Mean & Median & Std Dev.  & ($Q_{0.025},Q_{0.5},Q_{0.975}$)  \\ \hline
$\alpha_0$   & 1.42$\times 10^{-7}$ & 1.42$\times 10^{-7}$ & 7.11$\times 10^{-10}$ &
(1.41$\times 10^{-7}$,  1.42$\times 10^{-7}$, 1.43$\times 10^{-7}$) \\
\hline
$\alpha_1$ (day 12 ) & 4.62$\times 10^{-8}$ & 4.62 $\times 10^{-8}$ & 4.80 $\times 10^{-11}$ &
(4.62$\times 10^{-8}$, 4.62$\times 10^{-8}$4.63$\times 10^{-8}$) \\
\hline
$\alpha_2$ (day 35 ) & 5.99$\times 10^{-8}$ & 5.99 $\times 10^{-8}$ & 3.04$\times 10^{-10}$ &
(5.94$\times 10^{-8}$,  5.99$\times 10^{-8}$, 6.04 $\times 10^{-8}$) \\
\hline
$\alpha_3$ (day 48 ) & 4.38$\times 10^{-8}$ & 4.38$\times 10^{-8}$ &  8.86$\times 10^{-11}$ &
(4.36$\times 10^{-8}$, 4.38$\times 10^{-8}$, 4.39$\times 10^{-8}$) \\
\hline
$\alpha_4$ (day 60 ) & 4.32$\times 10^{-8}$ & 4.32 $\times 10^{-8}$ & 3.65 $\times 10^{-10}$ &
(4.25$\times 10^{-8}$, 4.32$\times 10^{-8}$, 4.37$\times 10^{-8}$) \\
\hline
$\alpha_5$ (day 71) & 3.63$\times 10^{-8}$ & 3.63 $\times 10^{-8}$ & 1.50 $\times 10^{-10}$ &
(3.61$\times 10^{-8}$, 3.63$\times 10^{-8}$, 3.66$\times 10^{-8}$) \\
\hline
$\alpha_6$ (day 78 ) &  4.96 $\times 10^{-10}$ & 5.63$\times 10^{-10}$ & 2.15$\times 10^{-10}$ &
(6.80$\times 10^{-11}$, 5.63$\times 10^{-10}$, 8.15$\times 10^{-10}$) \\
\hline
$\alpha_7$ (day 87 ) & 8.24 $\times 10^{-8}$ & 8.24 $\times 10^{-8}$ & 2.48$\times 10^{-10}$ &
(8.18$\times 10^{-8}$, 8.24$\times 10^{-8}$, 8.28$\times 10^{-8}$) \\
\hline
$\alpha_8$ (day 93 ) & 5.52 $\times 10^{-8}$ & 5.52$\times 10^{-8}$ & 1.27$\times 10^{-10}$ &
(5.50$\times 10^{-8}$, 5.52$\times 10^{-8}$, 5.55$\times 10^{-8}$) \\
\hline
$\alpha_{9}$ (day 104 ) &  9.63 $\times 10^{-12}$ & 6.87 $\times 10^{-12}$ &  8.95 $\times 10^{-12}$ &
(2.98$\times 10^{-13}$, 6.87$\times 10^{-12}$, 3.31$\times 10^{-11}$) \\
\hline
$\alpha_{10}$ (day 115 ) & 5.20 $\times 10^{-8}$ & 5.20 $\times 10^{-8}$ & 2.08$\times 10^{-10}$ &
(5.17$\times 10^{-8}$, 5.20 $\times 10^{-8}$, 5.24$\times 10^{-8}$) \\
\hline
$\alpha_{11}$ (day 136 ) & 5.79$\times 10^{-8}$ & 5.79 $\times 10^{-8}$ & 3.61$\times 10^{-11}$ &
(5.79$\times 10^{-8}$, 5.79$\times 10^{-8}$, 5.80$\times 10^{-8}$) \\
\hline
$\alpha_{12}$ (day 350) & 1.13$\times 10^{-8}$ & 1.13$\times 10^{-08}$ & 4.75 $\times 10^{-10}$ &
(1.05$\times 10^{-08}$, 1.13$\times 10^{-08}$, 1.23$\times 10^{-8}$) \\
\hline
$\alpha_{13}$ (day 355) & 3.35 $\times 10^{-8}$ & 3.35 $\times 10^{-8}$ & 9.35 $\times 10^{-11}$ &
(3.34$\times 10^{-8}$, 3.35$\times 10^{-8}$, 3.37$\times 10^{-8}$) \\
\hline
$\alpha_{14}$ (day 420 ) & 8.16 $\times 10^{-8}$ & 8.16 $\times 10^{-8}$ &  1.78 $\times 10^{-10}$ &
(8.13 $\times 10^{-8}$, 8.16 $\times 10^{-8}$, 8.20$\times 10^{-8}$) \\
\hline
$\beta^*$ (day 35) & 0.7888 & 0.7145 & 0.5703 & (0.02851, 0.62932, 4.64589 )  \\
\hline
$\beta$ &  0.05386 & 0.05383 & 0.00021 & (0.05350, 0.05383, 0.05420)  \\
\hline
$\gamma_0$ & 0.00846 & 8.48$\times 10^{-3}$ & 5.32$\times 10^{-05}$  & (0.00834, 0.00848, 0.00854)  \\
\hline
$\gamma_1$ & 0.01323 & 0.0132 & 5.96$\times 10^{-05}$ & (0.01310, 0.01324, 0.01334)  \\
\hline
$\gamma_2$ & 0.10092 & 0.10089 & 1.44$\times 10^{-04}$ & (0.10068, 0.10090, 0.10123)  \\
\hline
$\gamma_3$ & 0.09407 & 0.09406 & 7.13$\times 10^{-05}$ & (0.09392, 0.09406, 0.09420)  \\
\hline
$\gamma_4$ & 0.07121 & 0.07120 & 8.93 $\times 10^{-05}$ & (0.07106, 0.07120, 0.07142)  \\
\hline
$\gamma_5$ & 0.01979 & 0.01978 & 7.60 $\times 10^{-05}$ & (0.01966, 0.01978, 0.01993)  \\
\hline
$\gamma_6$ & 0.09309 & 0.09309 & 1.73 $\times 10^{-04}$ & (0.09278, 0.09309, 0.09343)  \\
\hline
$\rho$ & 0.00922 & 0.00922 & 3.35 $\times 10^{-06}$ & (0.00921, 0.00922, 0.00922 ) \\
\hline
$\zeta$ & 1.21 $\times 10^{-4}$ & 1.21$\times 10^{-4}$ & 3.53 $\times 10^{-07}$ & 1.21 $\times 10^{-04}$, 1.21 $\times 10^{-04}$, 1.22 $\times 10^{-04}$ ) \\
\hline
\end{tabular}
}
\end{table}


Since one of the aims of this paper is to understand the impact of the interventions enacted by the government on the transmission rate, simple hypothesis tests using contrasts were performed on the components of the transmission rate vector, $\bm{\alpha}$. Particularly, the sequential contrasts of $\alpha_1 -\alpha_0,~\alpha_2-\alpha_1,~\alpha_3-\alpha_2,~\alpha_4 -\alpha_3,\cdots, ~\alpha_{14} - \alpha_{13}$ were considered. These contrasts quantify the changes in the transmission rate due to the interventions, and this is of interest to the government and policymakers. Since these tests determine whether the intervention strategies implemented at different time points significantly impacted the dynamics of the pandemic, these will be more interesting to the policymakers. The contrasts needed for this analysis were obtained by subtracting posterior samples. From these sample differences, posterior estimates such as mean, median, standard deviation, quantiles, and the proportion of samples greater than 0, $P(>0)$ were obtained for the contrasts. The corresponding results are shown in Table~\ref{tbl:ConEst}.

\begin{table}[ht!]
\caption{Posterior Mean, Median, Standard Deviation, ($Q_{0.025},Q_{0.5},Q_{0.975}$) and proportion of samples larger than zero, $P(>0)$, for sequential contrasts across $\bm{\alpha}$, calculated based on 30,000 samples from the posterior distribution.}\label{tbl:ConEst}
\begin{center}
\scalebox{0.9}{
\begin{tabular}{|l c c c c c|} \hline
Contrast & Mean & Median & Std Dev. & ($Q_{0.025},Q_{0.5},Q_{0.975}$) & $P(>0)$ \\
\hline
$\alpha_1 - \alpha_0$ & -9.58$\times 10^{-8}$  & -9.57 $\times 10^{-8}$& 6.77$\times 10^{-10}$ & ( -9.70$\times 10^{-8}$,-9.57$\times 10^{-8}$, -9.48$\times 10^{-8}$) & 0.000 \\
\hline
$\alpha_2 - \alpha_1$ & 1.37$\times 10^{-8}$ & 1.37$\times 10^{-8}$& 3.35$\times 10^{-10}$ & ( 1.31$\times 10^{-8}$, 1.37$\times 10^{-8}$, 1.42$\times 10^{-8}$) & 1.000 \\
\hline
$\alpha_3 - \alpha_2$ & -1.61$\times 10^{-8}$ & -1.61$\times 10^{-8}$ & 2.49$\times 10^{-10}$ & ( -1.66$\times 10^{-8}$, -1.61$\times 10^{-8}$, -1.58$\times 10^{-8}$) & 0.000 \\
\hline
$\alpha_4 - \alpha_3$ & -5.50$\times 10^{-10}$ & -6.32$\times 10^{-10}$ & 4.31$\times 10^{-10}$ & ( -1.28$\times 10^{-9}$, -6.32$\times 10^{-10}$, 9.45 $\times 10^{-11}$) & 0.000 \\
\hline
$\alpha_5 - \alpha_4$ & -6.93$\times 10^{-9}$ & -6.92$\times 10^{-9}$ & 4.37$\times 10^{-10}$ & (-7.60$\times 10^{-9}$, -6.92$\times 10^{-9}$, -6.09$\times 10^{-9}$) & 0.000 \\
\hline
$\alpha_6 - \alpha_5$ & -3.58$\times 10^{-8}$ & -3.57$\times 10^{-8}$ &  2.82$\times 10^{-10}$ & (-3.65$\times 10^{-8}$, -3.57$\times 10^{-8}$, -3.54$\times 10^{-8}$) & 0.000 \\
\hline
$\alpha_7 - \alpha_6$ & 8.19$\times 10^{-8}$ & 8.19$\times 10^{-8}$ & 3.46$\times 10^{-10}$ & ( 8.12 $\times 10^{-8}$, 8.19$\times 10^{-8}$,  8.25$\times 10^{-8}$) & 1.000 \\
\hline
$\alpha_8 - \alpha_7$ & -2.71$\times 10^{-8}$ & -2.71 $\times 10^{-8}$ &2.86$\times 10^{-10}$ & ( -2.76$\times 10^{-8}$, -2.71$\times 10^{-8}$, -2.66$\times 10^{-8}$) & 0.000 \\
\hline
$\alpha_{9} - \alpha_8$ & -5.52$\times 10^{-8}$ & -5.52 $\times 10^{-8}$ & 1.28$\times 10^{-10}$ & ( -5.55$\times 10^{-8}$, -5.52$\times 10^{-8}$, -5.50 $\times 10^{-8}$) & 0.000 \\
\hline
$\alpha_{10} - \alpha_{9}$ &5.20$\times 10^{-8}$ & 5.20$\times 10^{-8}$ & 2.10$\times 10^{-10}$ & ( 5.17$\times 10^{-8}$, 5.20$\times 10^{-8}$, 5.24$\times 10^{-8}$) & 1.000 \\
\hline
$\alpha_{11} - \alpha_{10}$ & 5.92$\times 10^{-9}$ & 5.95$\times 10^{-9}$ & 1.85$\times 10^{-10}$ & ( 5.55$\times 10^{-9}$, 5.95$\times 10^{-9}$,  6.24$\times 10^{-9}$) & 1.000 \\
\hline
$\alpha_{12} - \alpha_{11}$ & -4.67$\times 10^{-8}$ & -4.67$\times 10^{-8}$ & 4.73$\times 10^{-10}$ & (-4.75$\times 10^{-8}$, -4.67$\times 10^{-8}$, -4.57$\times 10^{-8}$) & 0.000 \\
\hline
$\alpha_{13} - \alpha_{12}$ & 2.22$\times 10^{-8}$ & 2.22$\times 10^{-8}$ &4.71$\times 10^{-10}$ & ( 2.13$\times 10^{-8}$, 2.22$\times 10^{-8}$, 2.31$\times 10^{-8}$) & 1.000 \\
\hline
$\alpha_{14} - \alpha_{13}$ & 4.81$\times 10^{-8}$ & 4.81$\times 10^{-8}$ & 2.09$\times 10^{-10}$ & ( 4.77$\times 10^{-8}$, 4.81$\times 10^{-8}$, 4.85$\times 10^{-8}$) & 1.000 \\
\hline
\end{tabular}}
\end{center}
\end{table}

Notice that the intervention on day 12 ($\alpha_1 - \alpha_0$), reduced the transmission rate by approximately $9.58\times 10^{-8}$ and the proportion of samples above 0 was $0.000$ indicating a statistically significant change due to the intervention.  The intervention taken on day 35 ($\alpha_2-\alpha_1$) actually increased the transmission rate, which is evident from the positive contrast mean and the proportion of samples above 0 being 1.000. All intervention effects can be interpreted similarly. That is, interventions taken on days 48 $(\alpha_3-\alpha_2)$, 60 $(\alpha_4-\alpha_3)$, 71 $(\alpha_5-\alpha_4)$, 78 $(\alpha_6-\alpha_5)$, 93 $(\alpha_8-\alpha_7)$, 104 $(\alpha_9-\alpha_8)$ and 350 ($\alpha_{12}-\alpha_{11}$) were helpful in reducing the transmission rates whereas, the interventions implemented on days 87 $(\alpha_7-\alpha_6)$, 115 ($\alpha_{10}-\alpha_9$), 136 ($\alpha_{11}-\alpha_{10}$), 355 ($\alpha_{13}-\alpha_{12}$) and 420 ($\alpha_{14}-\alpha_{13}$) increased the transmission rate.

Connecting these contrasts to the interventions deployed by the Qatari government, the intervention measure taken by closing schools and universities on March 10, 2020 falls on day 48 ($\alpha_3-\alpha_2$) of the pandemic. Our result shows that this intervention was helpful in decreasing the transmission rate. Furthermore, the intervention deployed on March 21, 2020 by the Ministry of Municipality and Environment to close all parks and public beaches correspond to day 60 ($\alpha_4 -\alpha_3$) of the pandemic, the decision made by the Ministry of Commerce and Industry on April 1, 2020 to temporarily close down all restaurants, cafes and food trucks correspond to day 71 ($\alpha_5 -\alpha_4$), and the restrictions on the holy month of Ramadan correspond to day 93 ($\alpha_8 -\alpha_7$). These three interventions were helpful in the reduction of the transmission rates. 

The intervention deployed on June $4$, 2020, when the cabinet decided to allow four people inside a vehicle, exempting only families, and the intervention by the Ministry of Commerce and Industry (MoCL) to permit working hours for private sectors from 7am until 8pm fell on day 115 ($\alpha_{10}-\alpha_{9}$) of the pandemic. These two interventions were not effective as the transmission rate increased. 

This shows that some of the interventions by the Qatari government were effective in reducing the transmission rates of COVID-19, and if more interventions were taken, then we would see a tremendous overall reduction in the transmission rate. Furthermore, report shows that some restrictions were lifted on June $15$ (phase 1) through September, 2020 (phase 4), and there was not any major interventions since June $4$, 2020 which could be why the number of active infections increased between July 2020 and March 2021.  Furthermore, results in Table ~\ref{tbl:ConEst} shows that if an intervention took place on day 350 ($\alpha_{12}-\alpha_{11}$), it would have reduced the transmission rate.


We also conducted simple hypothesis tests on sequential contrasts of the components of $\bm{\gamma}$, namely, $\gamma_1 -\gamma_0,~\gamma_2-\gamma_1,~\gamma_3-\gamma_2,~\gamma_4 -\gamma_3,~\gamma_5 -\gamma_4, ~\gamma_6 - \gamma_5$ to look for significant changes in the recovery rate due to interventions. These tests allow us to verify if the interventions performed significantly impacted the dynmics of the pandemic. The estimated contrasts are obtained similar to the transmission rate contrasts. Table \ref{tbl:ConGk} shows the results for the recovery rate contrasts. 

\begin{table}[H]
\caption{Posterior Mean, Median, Standard Deviation, ($Q_{0.025},Q_{0.5},Q_{0.975}$) and proportion of samples larger than zero, $P(>0)$, for sequential contrasts across $\bm{\gamma}$, calculated based on 30,000 samples from the posterior distribution.}\label{tbl:ConGk}
\begin{center}
\scalebox{0.9}{
\begin{tabular}{|l c c c c c|} \hline
Contrast & Mean & Median & Std Dev. & ($Q_{0.025},Q_{0.5},Q_{0.975}$) & $P(>0)$ \\
\hline
$\gamma_1 - \gamma_0$ & 4.71$\times 10^{-9}$  & 4.77 $\times 10^{-9}$ & 1.55$\times 10^{-9}$ & ( 4.31$\times 10^{-9}$, 4.77 $\times 10^{-9}$, 4.90 $\times 10^{-9}$) & 1.000 \\
\hline
   
$\gamma_2 - \gamma_1$ & 8.77$\times 10^{-8}$ & 8.77$\times 10^{-8}$& 1.16$\times 10^{-9}$ & (8.75$\times 10^{-8}$, 8.77$\times 10^{-8}$, 8.80 $\times 10^{-8}$) & 1.000 \\
\hline
$\gamma_3 - \gamma_2$ & -6.95$\times 10^{-9}$ & -6.95$\times 10^{-9}$ & 1.05$\times 10^{-9}$ & (-7.19$\times 10^{-9}$, -6.95 $\times 10^{-9}$, -6.78 $\times 10^{-9}$) & 0.000 \\
\hline
$\gamma_4 - \gamma_3$ &-2.29$\times 10^{-8}$ & -2.28$\times 10^{-8}$ & 7.52$\times 10^{-5}$ & ( -2.30 $\times 10^{-8}$, -2.28 $\times 10^{-8}$, -2.27$\times 10^{-8}$) & 0.000 \\
\hline
$\gamma_5 -\gamma_4$ & -5.14$\times 10^{-9}$ &-5.14$\times 10^{-8}$ & 1.53$\times 10^{-9}$ & (-5.16$\times 10^{-8}$, -5.14 $\times 10^{-8}$, -5.10 $\times 10^{-8}$) & 0.000 \\
\hline
$\gamma_6 - \gamma_5$ & 7.34$\times 10^{-9}$ & 7.33$\times 10^{-8}$ & 2.34$\times 10^{-10}$ & (7.29 $\times 10^{-8}$, 7.33 $\times 10^{-8}$, 7.38$\times 10^{-8}$) & 1.000 \\
\hline
\end{tabular}}
\end{center}
\end{table}

To assess the fit of the model the posterior predictive distribution was used and is given by:
\begin{eqnarray}
\pi( I_{new}(t), R_{I_{new}}(t), D_{new}(t),  V_{new}(t) | \mathbf{D} ) &=& \int L( I_{new}(t), R_{I_{new}}(t), D_{new},  V_{new}(t)|\bm{ \alpha }, \beta^{*}, \beta,\bm{\gamma},\zeta) \cr
&\times& \pi \left( \bm{ \alpha }, \beta^{*}, \beta, \bm{\gamma}, \zeta | \mathbf{D} \right) d \bm{ \alpha }d\beta^{*} d\beta d \bm{\gamma} d \zeta .
\end{eqnarray}

\noindent Using the posterior samples, 30,000 samples were generated from the posterior predictive distribution.  At each time $t$, the posterior median, 0.025 and 0.975 quantiles were also obtained to form a posterior predictive interval.

\begin{figure}
 \begin{subfigure}{0.5\textwidth}
     \includegraphics[width=\textwidth]{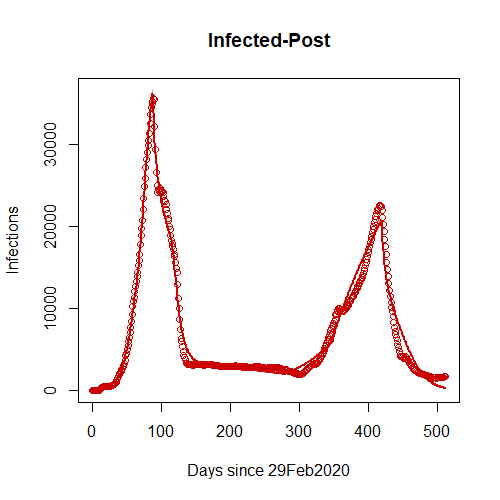}
     \caption{Number of active infections }
     \label{fig:a2}
 \end{subfigure}
 \hfill
 \begin{subfigure}{0.5\textwidth}
     \includegraphics[width=\textwidth]{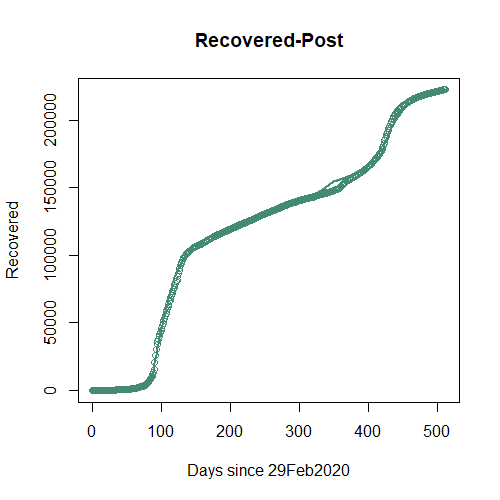}
     \caption{Cumulative Number of recovered individuals}
     \label{fig:b2}
 \end{subfigure}
 
 \medskip
 \begin{subfigure}{0.5\textwidth}
     \includegraphics[width=\textwidth]{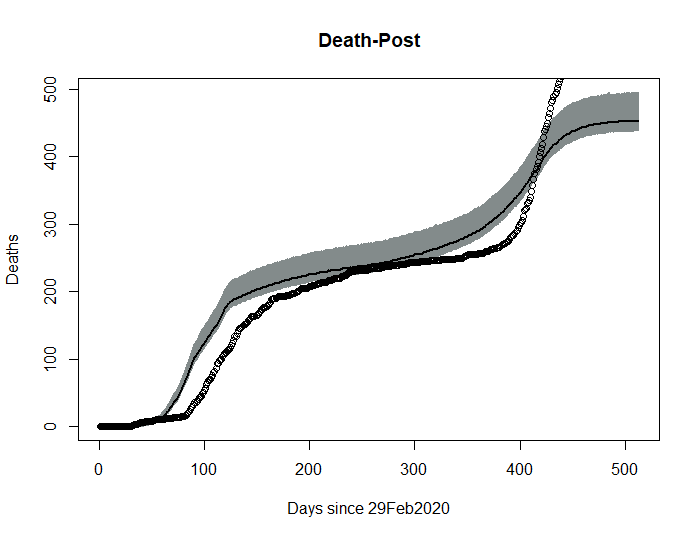}
     \caption{Cumulative Number of deaths}
     \label{fig:c2}
 \end{subfigure}
 \hfill
 \begin{subfigure}{0.5\textwidth}
     \includegraphics[width=\textwidth]{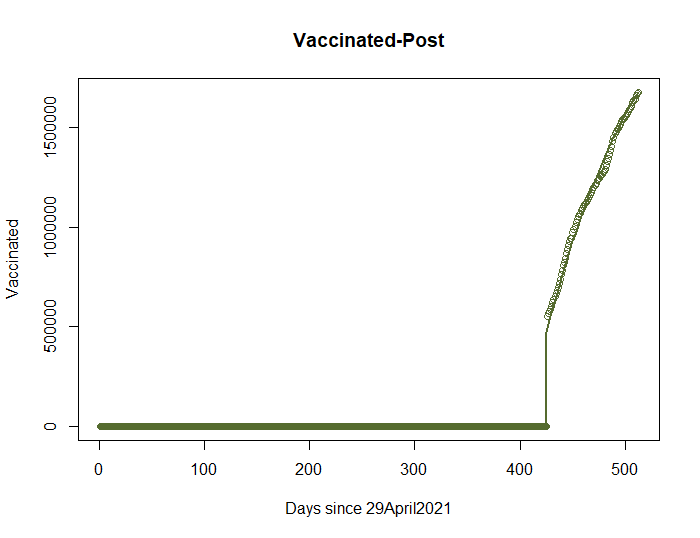}
     \caption{Cumulative number of vaccinated individuals}
     \label{fig:d2}
 \end{subfigure}

 \caption{Plots of fits for (a) Active Infections, (b) Recovered, (c) Deaths (February 29, 2020 - October 13, 2021) and (d) Vaccinated (April 29, 2021 - October 13, 2021) for the State of Qatar. for the days since 29 February 2020 until 13 October 2021. Posterior predictive bands are the 0.025 and 0.975 quantiles calculated using 30,000 samples from the posterior predictive distribution.}
 \label{fig:PostData}
 \end{figure}

Figure~\ref{fig:PostData} shows the model fits for Active Infections, Recovered, Deaths and Vaccinated with posterior predictive bands.  Notice that, the model does quite well at fitting the dynamics of the Active Infections including the jumps at days 95 and 420, and captures  the plateau and the exponential growth after the plateau as well.  The model also fits the cumulative counts of recovered, deaths and vaccinated individuals quite well. To assess the explained variance, a pseudo-$R^2$ \citep{boone2021monitoring} was formed using the posterior median at each time point.  This resulted in a pseudo-$R^2$ of 0.9995 which indicates the fitted model explains approximately 99.9\% of the variance in the data. In order to assess the validity and robustness of our model, we performed a validation test by fitting our model to COVID-19 data from Nigeria. Our model also fit this data reasonably well with a pseudo-$R^2 \approx 0.801$.

\subsection{Behavior of $\mathfrak{R_e} (t)$ over time}

\begin{figure}[ht!]
\begin{center}
\includegraphics[width = 0.7\textwidth]{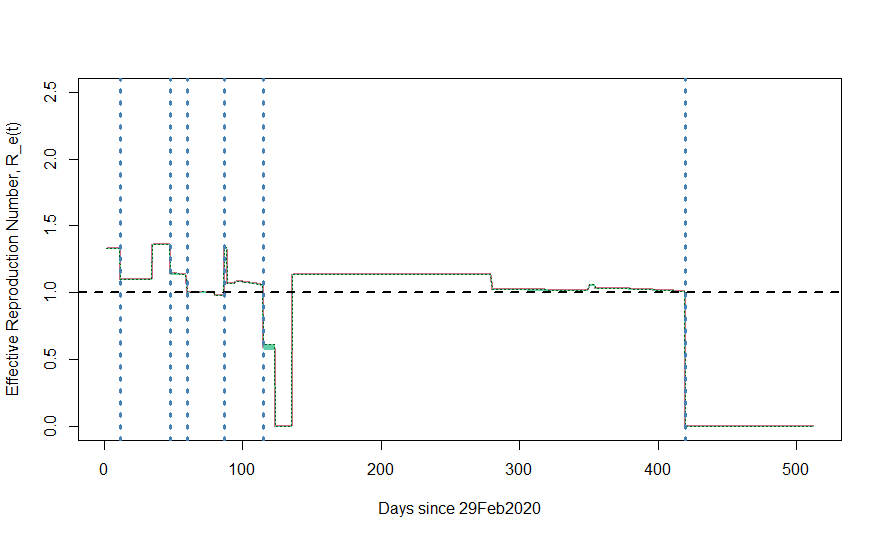} 
\caption{Plots of Time Varying Effective Reproduction Number for the State of Qatar for the days since 29 February 2020 until until 13 October 2021 with the 0.025 and 0.975 Quantiles . \label{fig:RepNum}}
\end{center}
\end{figure}

Figure \ref{fig:RepNum} shows the estimated time varying effective reproduction number, $\mathfrak{R_e} (t)$, along with 95\% confidence bands (colored green). Until about day 90, the entire band is above $1$, showing that the system was in a pandemic state at the beginning of the study period. The estimated $\mathfrak{R_e} (t)$ in this period is about $1.4$, which indicates that the number of secondary infections was high and more people could get infected since the population was well mixed. For the time period when the band is close to or contains 1, it is difficult to conclude whether the system was in a pandemic state or not. But after the introduction of vaccines on day 420, $\mathfrak{R_e} (t)$ goes below 1, showing that the system was then in the endemic or declining state. 

Additionally, we can visualize the impact of the intervention strategies by the Qatari government using the time varying reproduction number. With an intervention on day $48$, $\mathfrak{R_e}(t)$ declines from about $1.4$ to about $1.1$. This shows that the intervention on $48$ was helpful in reducing the spread of secondary infections. Similarly, the intervention implemented on day $60$ was also effective in controlling the spread of COVID-19, as reflected in the drop of $\mathfrak{R_e}(t)$ from about $1.1$ to $1$. These conclusions consistent with the results in Table \ref{tbl:ConEst}. After the introduction of vaccination on day 420, $\mathfrak{R_e}(t)$ drops drastically to about $0.05$, showing the importance of getting vaccinated. This is also evident from Figure \ref{fig:PostData}(b), where the number of infected dropped drastically when vaccination was introduced. Between days 126 and 420, we could not decide whether the system was in the pandemic state, which reflects the government's decision to lift restrictions on June 15, 2020 (day 126) and to not implement any major intervention measures until April 29, 2021 (day 420), when vaccines were introduced in Qatar. 

\subsection{Effect of Vaccination on the Number of Deaths}
\begin{figure}[ht!]
\begin{center}
\includegraphics[width = 0.7\textwidth]{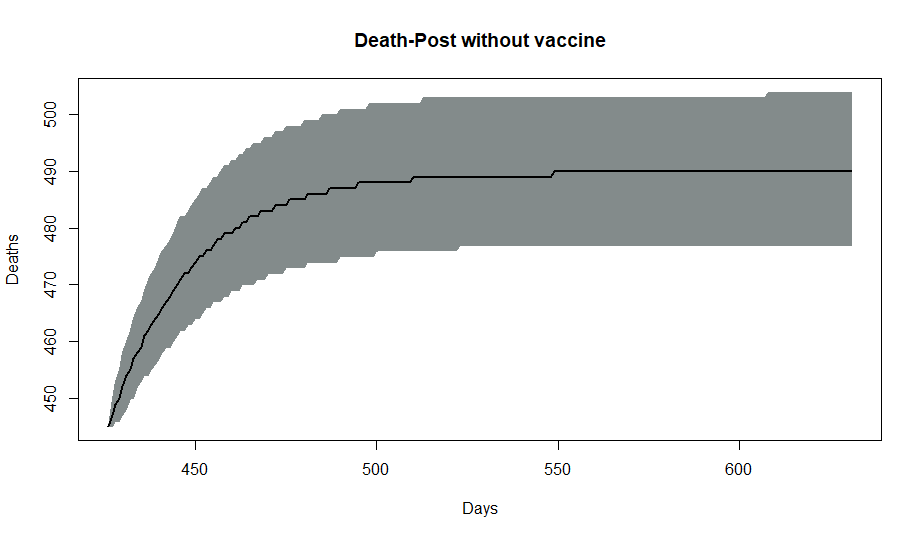} 
\caption{Projected number of deaths without introduction of the vaccine, along with 95\% prediction intervals. \label{fig:proj}}
\end{center}
\end{figure}

In addition to quantifying the effect of vaccination on the number of secondary infections and the death rate, the proposed SEIRDV model also allows for the quantification of how many lives were saved due to the vaccine, that is, how many excess deaths were prevented due to vaccination over the study time frame. This question is crucial because it allows decision makers to understand the impact of vaccination. According to \cite{mathieu2021death}, death is a key metric that accurately shows how effective vaccines are against the most severe form of diseases. Figure \ref{fig:proj} shows the cumulative deaths that would have been observed had the vaccine not been deployed and the lock down conditions present at the time remained in effect during the remainder of the study period.  The associated 95\% prediction intervals are given as well. The prediction interval at the last time point shows that between $32$ and $59$ lives were saved on that day because of vaccination. Also, note that cumulative deaths would have plateaued even with strict lockdown policies. Thus, this is evidence that vaccination prevented excess deaths and were instrumental in the relaxation of the restrictions, allowing people to return to their normal lives. 

\section{Discussion}\label{sec:Discussion}
\noindent This work provides a novel extension of the Susceptible, Exposed, Infected, Recovered, Death (SEIRD) model to the SEIRDV model by adding a Vaccination compartment and incorporating interventions to help understand the impact of government policies on disease transmission rates. Additionally the impact of vaccination is also studied using time varying effective reproduction number. All inference are made under the Bayesian framework. The model is able to treat the Susceptible and Exposed compartments as latent variables, since no data is observed about them other than approximate initial values. The model is implemented on COVID-19 dataset for the State of Qatar. The model fits the data quite well with a pseudo-$R^2 \approx 0.9997$. Results show that the strict interventions deployed by the government of Qatar such as closure of schools, recreational centers, restaurants, working from home, travel bans, and prevention of social gathering reduced the transmission rate while the non-strict interventions could not reduce the transmission rate. The estimated time-varying effective reproduction number in Figure \ref{fig:RepNum} shows that the system was strictly in a declining state after the deployment of vaccines (pharmaceutical intervention). 

A validation test was performed using a different dataset (Nigeria dataset) to check for the robustness of the model, and the result shows that the model fits this external dataset quite well with a pseudo $R^2 \approx 0.801$.


The modeling framework is quite flexible for modeling the COVID-19 data as it easily incorporates external interventions into the system by varying transmission and recovery rates over the study period and can also quantify the impact of the external interventions using sequential contrasts of these rates. The sequential contrasts are used to make inference about the intervention effects. For the State of Qatar, strict/severe interventions such as the closure of schools (day 48), parks, restaurants, bars, and travel bans (days 60 and 71) were effective in reducing the transmission rate, while non-strict/liberal interventions increased the transmission rate. Using the proposed model, we demonstrated the huge impact of vaccination on the effective reproduction number, compared to other intervention measures that were deployed when the pandemic was at its peak. Figure \ref{fig:PostData} also shows that the number of infected increased drastically when there were no interventions placed by the government. This is not surprising because intervention measures are limited and transient, especially the strict ones. Therefore, vaccinations seem to be the metric that is effective and sustainable in reducing the number of infected people, especially now that social distancing might not be feasible. Our results in Tables \ref{fig:RepNum} and  \ref{fig:PostData} clearly show that the more people become vaccinated, the more drastic the number of active infections would drop. Although most people had the notion that vaccination is not effective since we still have many cases (infected), our model shows that the vaccine is indeed effective in reducing the number of secondary cases, thereby checking the intensity of the pandemic.

After a thorough scan of the literature, we were unable to find a study that quantified the number of excess deaths prevented as a result of vaccine implementation. The work here shows how one can quantify this using a novel mathematical approach using public heath data and a counter-factual scenario. This can be an effective tools for decision makers who are faced with making difficult decisions about public health policies in the event of a pandemic. 


\section*{Acknowledgements}

\noindent The authors would like to thank Virginia Commonwealth University in Qatar and Qatar Foundation for supporting this work through the VCU Qatar Mathematical Data Sciences Lab.

\bibliographystyle{jds}
\bibliography{JDSbib}

\begin{thebibliography}{25}
\providecommand{\natexlab}[1]{#1}

\bibitem[{Albert(2009)}]{albert2009introduction}
Albert J (2009).
\newblock An introduction to r.
\newblock In: \emph{Bayesian Computation with R}, 1--17. Springer.

\bibitem[{Bayes(1763)}]{bayes1763lii}
Bayes T (1763).
\newblock Lii. an essay towards solving a problem in the doctrine of chances.
  by the late rev. mr. bayes, frs communicated by mr. price, in a letter to
  john canton, amfr s.
\newblock \emph{Philosophical transactions of the Royal Society of London},
  (53): 370--418.

\bibitem[{Berger(1985)}]{berger1985prior}
Berger JO (1985).
\newblock Prior information and subjective probability.
\newblock In: \emph{Statistical Decision Theory and Bayesian Analysis},
  74--117. Springer.

\bibitem[{Bertsimas et~al.(2020)Bertsimas, Ivanhoe, Jacquillat, Li, Previero,
  Lami et~al.}]{Bertsimas2020optimizing}
Bertsimas D, Ivanhoe J, Jacquillat A, Li M, Previero A, Lami OS, et~al. (2020).
\newblock Optimizing vaccine allocation to combat the covid-19 pandemic.
\newblock \emph{medRxiv}.

\bibitem[{Boone et~al.(2021)Boone, Abdel-Salam, Sahoo, Ghanam, Chen, and
  Hanif}]{boone2021monitoring}
Boone EL, Abdel-Salam ASG, Sahoo I, Ghanam R, Chen X, Hanif A (2021).
\newblock Monitoring seird model parameters using mewma for the covid-19
  pandemic with application to the state of qatar.
\newblock \emph{Journal of Applied Statistics}, 1--16.

\bibitem[{Casella and Berger(2021)}]{casella2021statistical}
Casella G, Berger RL (2021).
\newblock \emph{Statistical inference}.
\newblock Cengage Learning.

\bibitem[{Chinazzi et~al.(2020)Chinazzi, Davis, Ajelli, Gioannini, Litvinova,
  Merler et~al.}]{chinazzi2020effect}
Chinazzi M, Davis JT, Ajelli M, Gioannini C, Litvinova M, Merler S, et~al.
  (2020).
\newblock The effect of travel restrictions on the spread of the 2019 novel
  coronavirus (covid-19) outbreak.
\newblock \emph{Science}, 368(6489): 395--400.

\bibitem[{Dirac(1958)}]{dirac1958general}
Dirac P (1958).
\newblock General principles of quantum mechanics.
\newblock \emph{International series of monographs on physic. Oxford, Clarendon
  Press, 4th edition edition}.

\bibitem[{Gelman et~al.(1995)Gelman, Carlin, Stern, and
  Rubin}]{gelman1995bayesian}
Gelman A, Carlin JB, Stern HS, Rubin DB (1995).
\newblock \emph{Bayesian data analysis}.
\newblock Chapman and Hall/CRC.

\bibitem[{Ghanam et~al.(2021)Ghanam, Boone, and Abdel-Salam}]{ghanam2021seird}
Ghanam R, Boone EL, Abdel-Salam ASG (2021).
\newblock Seird model for qatar covid-19 outbreak: A case study.
\newblock \emph{Letters in Biomathematics}, 8(1).

\bibitem[{Ghostine et~al.(2021)Ghostine, Gharamti, Hassrouny, and
  Hoteit}]{Ghostine2021extended}
Ghostine R, Gharamti M, Hassrouny S, Hoteit I (2021).
\newblock An extended seir model with vaccination for forecasting the covid-19
  pandemic in saudi arabia using an ensemble kalman filter.
\newblock \emph{Mathematics}, 9(6): 636.

\bibitem[{Gilks et~al.(1995)Gilks, Richardson, and
  Spiegelhalter}]{gilks1995markov}
Gilks WR, Richardson S, Spiegelhalter D (1995).
\newblock \emph{Markov chain Monte Carlo in practice}.
\newblock CRC press.

\bibitem[{Giuliani et~al.(2020)Giuliani, Dickson, Espa, and
  Santi}]{giuliani2020modelling}
Giuliani D, Dickson MM, Espa G, Santi F (2020).
\newblock Modelling and predicting the spatio-temporal spread of covid-19 in
  italy.
\newblock \emph{BMC infectious diseases}, 20(1): 1--10.

\bibitem[{Kermack and McKendrick(1927)}]{kermack1927contribution}
Kermack WO, McKendrick AG (1927).
\newblock A contribution to the mathematical theory of epidemics.
\newblock \emph{Proceedings of the royal society of london. Series A,
  Containing papers of a mathematical and physical character}, 115(772):
  700--721.

\bibitem[{Koufi et~al.(2020)Koufi, Bennar, and Yousfi}]{Koufi2020dynamics}
Koufi AE, Bennar A, Yousfi N (2020).
\newblock Dynamics of a stochastic sirs epidemic model with regime switching
  and specific functional response.
\newblock \emph{Discrete Dynamics in Nature and Society}, 2020.

\bibitem[{Mathieu and Roser(2021)}]{mathieu2021death}
Mathieu E, Roser M (2021).
\newblock How do death rates from covid-19 differ between people who are
  vaccinated and those who are not.
\newblock \emph{Our Wordl in data. Recuperado de: https://ourworldindata.
  org/covid-deaths-by-vaccination}.

\bibitem[{Mercer et~al.(2011)Mercer, Glass, and Becker}]{Mercer2011effective}
Mercer GN, Glass K, Becker NG (2011).
\newblock Effective reproduction numbers are commonly overestimated early in a
  disease outbreak.
\newblock \emph{Statistics in medicine}, 30(9): 984--994.

\bibitem[{Miller(2020)}]{miller20202019}
Miller M (2020).
\newblock 2019 novel coronavirus covid-19 (2019-ncov) data repository: Johns
  hopkins university center for systems science and engineering.
\newblock \emph{Bulletin-Association of Canadian Map Libraries and Archives
  (ACMLA)}, (164): 47--51.

\bibitem[{Rey et~al.(2021)Rey, Hammad, and Saberi}]{Rey2021vaccine}
Rey D, Hammad AW, Saberi M (2021).
\newblock Vaccine allocation policy optimization and budget sharing mechanism
  using thompson sampling.
\newblock \emph{arXiv preprint arXiv:2109.10004}.

\bibitem[{Rezabakhsh et~al.(2020)Rezabakhsh, Ala, and
  Khodaei}]{rezabakhsh2020novel}
Rezabakhsh A, Ala A, Khodaei SH (2020).
\newblock Novel coronavirus (covid-19): a new emerging pandemic threat.
\newblock \emph{Journal of Research in Clinical Medicine}, 8(1): 5--5.

\bibitem[{Ridenhour et~al.(2014)Ridenhour, Kowalik, and
  Shay}]{Ridenhour2014unraveling}
Ridenhour B, Kowalik JM, Shay DK (2014).
\newblock Unraveling r 0: Considerations for public health applications.
\newblock \emph{American journal of public health}, 104(2): e32--e41.

\bibitem[{Van~den Driessche and Watmough(2002)}]{van2002reproduction}
Van~den Driessche P, Watmough J (2002).
\newblock Reproduction numbers and sub-threshold endemic equilibria for
  compartmental models of disease transmission.
\newblock \emph{Mathematical biosciences}, 180(1-2): 29--48.

\bibitem[{Wackerly et~al.(2014)Wackerly, Mendenhall, and
  Scheaffer}]{wackerly2014mathematical}
Wackerly D, Mendenhall W, Scheaffer RL (2014).
\newblock \emph{Mathematical statistics with applications}.
\newblock Cengage Learning.

\bibitem[{Wintachai and Prathom(2021)}]{Wintachai2021stability}
Wintachai P, Prathom K (2021).
\newblock Stability analysis of seir model related to efficiency of vaccines
  for covid-19 situation.
\newblock \emph{Heliyon}, 7(4): e06812.

\bibitem[{Wu et~al.(2020)Wu, Zhao, Yu, Chen, Wang, Song et~al.}]{wu2020new}
Wu F, Zhao S, Yu B, Chen YM, Wang W, Song ZG, et~al. (2020).
\newblock A new coronavirus associated with human respiratory disease in china.
\newblock \emph{Nature}, 579(7798): 265--269.

\end{thebibliography}

\end{document}